\documentclass[preprint,showpacs,preprintnumbers,amsmath,amssymb]{revtex4}

\usepackage{graphicx}
\usepackage{dcolumn}
\usepackage{bm}
\usepackage{subfigure}
\usepackage{color}

\begin{document}


\title{Detecting the superfluid critical momentum of Bose gases in optical lattices through dipole oscillations
}

\author{Takuya Saito$^1$}
 \email{takuya.saito.880217@gmail.com}
\author{Ippei Danshita$^2$}%
 \email{danshita@riken.jp}
 \author{Takeshi Ozaki$^1$}
\author{Tetsuro Nikuni$^1$}
\affiliation{$^1$Department of Physics , Faculty of Science, Tokyo University of Science, 1-3 Kagurazaka, Shinjuku, Tokyo 162-8601, Japan}
\affiliation{
$^2$Computational Condensed Matter Physics Laboratory, RIKEN, 2-1 Hirosawa, Wako, Saitama 351-0198, Japan
}%

\date{\today}

\begin{abstract}
We study stability of superflow of Bose gases in optical lattices by analyzing the Bose-Hubbard model within the Gutzwiller mean-field approximation. We calculate the excitation spectra of the homogeneous Bose-Hubbard model at unit filling to determine the critical momenta for the Landau and dynamical instabilities. These two critical momenta are shown to approach each other when the on-site interaction increases towards the Mott transition point. In order to make a direct connection with realistic experiments, we next take into account a parabolic trapping potential and compute the real-time dynamics of dipole oscillations induced by suddenly displacing the trap center. We consider the following two cases: standard softcore bosons, whose interparticle interactions include the on-site one only, and hardcore bosons with long-range dipole-dipole interactions. For both cases, we show that the dipole oscillation is significantly damped when the maximum local momentum exceeds a certain threshold, which quantitatively agrees with the critical momentum for the dynamical instability in the homogeneous system. 
In the case of dipolar hardcore bosons, the dynamical instability of dipole oscillations  leads to the formation of checkerboard density waves in the superfluid phase near the boundary to the supersolid phase.
\end{abstract}

\pacs{Valid PACS appear here}
\maketitle

\section{Introduction\label{sec.1}}
Since the observation of the quantum phase transition from superfluid (SF) to Mott insulator (MI)~\cite{nature2002}, systems of ultracold gases confined in optical lattices have provided an ideal venue for studying strong correlation physics~\cite{RMP80_885}. Many interesting properties of strongly correlated ultracold gases have been revealed in recent experiments, including excitation spectra~\cite{PRL92_130403, PRL102_155301, NP6_56, PRL106_205303}, quantum criticality~\cite{science2012}, and particle transport~\cite{PRL94_120403, PRL99_150604, nature2008}. 

In particular, transport properties have attracted much interest, in connection with the SF critical momentum above which superflow breaks down. In experiments with ultracold gases in optical lattices, transport is investigated by using a moving optical lattice~\cite{PRL99_150604, PRL93_140406, PRA72_013603}  or suddenly displacing the parabolic trapping potential to induce a dipole oscillation~\cite{PRL94_120403, nature2008, PRL86_4447}. In the weakly interacting regime, where the Gross-Pitaevskii (GP) mean-field approximation is valid, it has been shown that the critical momentum in a trapped system at low temperatures agrees quantitatively with that for the dynamical instability (DI) in the homogeneous lattice system~\cite{PRL86_4447, PRL93_140406, PRA72_013603, PRA64_061603, PRA70_043625}, regardless of whether one uses the trap displacement or the moving optical lattice. In the intermediate and strongly-interacting regimes, while the agreement in the critical momenta of the trapped and untrapped systems has been obtained when the moving optical lattice is used~\cite{PRL99_150604, PRL95_020402, PRA71_063613}, it remains unclear whether this is also the case for the trap displacement.

In addition to optical lattices, new possibilities for the study of strong correlation physics have been opened up by the creation of ultracold atomic gases with strong magnetic dipole-dipole interactions~\cite{PRL94_160401, PRL107_190401} and gases of polar molecules~\cite{science2008, PRL105_203001}. It has been predicted that when a dipolar Bose gas is loaded on an optical lattice, there exist supersolid (SS) phases that possess both diagonal (crystalline) and off-diagonal (superfluid) long-range order~\cite{PRL88_170406, EPL72_162, PRL98_260405, PRL103_225301, PRL104_125301, PRL104_125302, PRA82_043625, JPSJ2011}. Danshita and Yamamoto have studied the critical momenta of dipolar Bose gases in a two-dimensional (2D) optical lattices and suggested that some properties of the critical momenta can be used to identify supersolid phases~\cite{PRA82_043625}. More specifically, the critical momenta of the SS phases are finite in contrast to the insulating phases, such as MI and density wave phases, and significantly smaller than that for the SF phase. Since the analyses in Ref.~\cite{PRA82_043625} have been done in a homogeneous lattice system, it is important to investigate the critical momenta of dipolar Bose gases in the presence of a parabolic trap in order to make the suggestion more convincing.

In this paper, using the Gutzwiller approximation, we study superfluid transport of Bose gases in optical lattices with and without the dipole-dipole interactions. First, we analyze the critical momenta of Bose gases without the dipole-dipole interactions at unit filling in homogeneous 1D, 2D, and 3D optical lattices by calculating the excitation spectra. We locate the boundaries to the Landau instability (LI) and the DI. Secondly, solving the equation of motion numerically, we simulate the dynamics of dipole oscillations of Bose gases confined in a parabolic potential both in the absence and the presence of the dipole-dipole interactions. For both cases, we find significant damping of the dipole oscillations when the maximum local momentum exceeds a certain critical value, and that the critical value coincides with the critical momentum for the DI in the homogeneous lattice systems. We also find a parameter region in which the dipole-oscillation mode is resonantly coupled with the breathing mode. When the dipole-dipole interactions are present and the system is in the SF state close to the SS phase, we show that the DI of a dipole oscillation is followed by the formation of checkerboard density waves, as predicted in Ref.~\cite{PRA82_043625}.

This paper is organized as follows. 
In Sec.~\ref{sec.2}, we introduce the Bose-Hubbard Hamiltonian and our formulation of the problem based on the Gutzwiller mean-field approximation.
In Sec.~\ref{sec.3}, we calculate the critical momenta at unit filling in homogeneous 1D, 2D, and 3D optical lattices from the excitation spectra. 
In Sec.~\ref{sec.4}, we calculate the dipole oscillations to determine the critical momenta in the presence of a parabolic trapping potential.
In Sec.~\ref{sec.5}, we analyze the dipole oscillations of dipolar hardcore bosons.
In Sec.~\ref{sec.6}, we summarize our results.

\section{Model and formulation\label{sec.2}}
\subsection{Bose-Hubbard Model}
We consider a system of ultracold Bose gases confined in deep optical lattices combined with a parabolic trap. 
This system is well described by the Bose-Hubbard model~\cite{PRB40_546, PRL81_3108},
\begin{eqnarray}
\hat{H}=-J\sum_{\bf j}\sum_{\alpha = 1}^{d}(\hat{a}^{\dag}_{\bf j}\hat{a}_{{\bf j} + {\bf e}_{\alpha}}+{ \rm H.c.})
+\frac{U}{2}\sum_{\bf j} \hat{n}_{\bf j}(\hat{n}_{\bf j}-1) 
+ \sum_{\bf j} (\epsilon_{\bf j} - \mu)  \hat{n}_{\bf j},
\label{eq_H}
\end{eqnarray}
where $\epsilon_{\bf j} = \Omega |a {\bf j}-{\bf r}_0(t)|^2$. $a$ and $d$ represent the lattice constant and the spatial dimension. The vectors ${\bf j}$ and ${\bf e}_{\alpha}$ denote the site index and a unit vector in the direction $\alpha$, where the directions $\alpha = 1, 2,$ and $3$ mean the $x$, $y$, and $z$ directions. $\hat{a}_{\bf j}$ ($\hat{a}^{\dag}_{\bf j}$) is the annihilation (creation) operator and $\hat{n}_{\bf j}=\hat{a}^{\dag}_{\bf j}\hat{a}_{\bf j}$ is the number operator at the site ${\bf j}$. $J$, $U$, $\Omega$, and $\mu$ are the hopping amplitude,  the on-site interaction parameter, the curvature of the parabolic trap, and the chemical potential, respectively. We fix the trap curvature to be $\Omega/J=0.01$ when we consider dipole oscillations of Bose gases in optical lattices in the presence of a parabolic trap. We assume that dipole oscillations are induced by displacing the trap center in the $x$ direction as $\bm{r}_0(t)= D a\theta (t)\hat{\bm{e}}_1$ , where $\theta (t)$ is the step function. We control the momentum of Bose gases by changing the displacement $Da$.  Henceforth, we adopt the units in which $J=\hbar=a=1$ except in the figures and their captions.

\subsection{Gutzwiller approximation}
In order to calculate the ground states, the excitation spectra, and the real-time dynamics of Bose gases in optical lattices, we use the Gutzwiller approximation, in which the many-body wave-function is assumed to be a single product of local states as~\cite{PRB44_10328}, 
\begin{eqnarray}
|\Psi_G\rangle=\prod_{\bf j} \sum_{n}f_{{\bf j},n}(t)|n\rangle_{\bf j},\label{eq_Gutzwiller}
\end{eqnarray}
where $|n\rangle_{\bf j}$ represents the local Fock state at the site ${\bf j}$ and the normalization condition for coefficient $f_{{\bf j}, n}$ is $\sum_n|f_{{\bf j}, n}|^2=1$.
Minimizing the effective action, $\displaystyle\int dt \left\langle \Psi_G\left| i\frac{d}{dt}-\hat{H}\right|\Psi_G\right\rangle$, with respect to $f_{{\bf j}, n}^{\ast}$, one can derive the equation of motion for $f_{{\bf j}, n}$,
\begin{eqnarray}
i\frac{df_{{\bf j},n}(t)}{dt}&=&-\sum_{\alpha=1}^{d} 
\left[ \sqrt{n} f_{{\bf j}, n-1} (\Phi_{{\bf j} - {\bf e}_{\alpha}} + \Phi_{{\bf j} + {\bf e}_{\alpha}}) 
+\sqrt{n+1}f_{{\bf j}, n+1} (\Phi_{{\bf j} - {\bf e}_{\alpha}}^{\ast} + \Phi_{{\bf j} + {\bf e}_{\alpha}}^{\ast}) \right]
\nonumber \\
&&+\left[ \frac{U}{2}n(n-1) + (\epsilon_{\bf j} - \mu)n \right]f_{{\bf j}, n}
\label{eq_df/dt},
\end{eqnarray} 
where $\Phi_j \equiv\langle \Psi_G|\hat{a}_j|\Psi_G\rangle = \sum_n \sqrt{n}f_{{\bf j},n-1}^{\ast}f_{{\bf j},n}$ is the superfluid order parameter.
While Eq.~(\ref{eq_df/dt}) obviously describes the real-time dynamics, it also allows us to calculate the ground states by solving it in imaginary-time~\cite{PRA76_023606}. 
A steady solution takes the form of $f_{{\bf j},n}(t)=\tilde{f}_{{\bf j},n}e^{-i\tilde{\omega}_{\bf j} t}$, where $\tilde{f}_{{\bf j},n}$ is time-independent and the phase factor
$\tilde{\omega}_{\bf j}$ is given by
\begin{eqnarray}
\tilde{\omega}_{\bf j}=
-\sum_{\alpha = 1}^{d} \left[ 
(\tilde{\Phi}_{{\bf j}-{\bf e}_{\alpha}} + \tilde{\Phi}_{{\bf j}+{\bf e}_{\alpha}})\tilde{\Phi}^{\ast}_{\bf j}
+(\tilde{\Phi}^{\ast}_{{\bf j}-{\bf e}_{\alpha}}+\tilde{\Phi}^{\ast}_{{\bf j}+{\bf e}_{\alpha}})\tilde{\Phi}_{\bf j}
\right]
+\sum_n\left[\frac{U}{2}(n-1)-\mu_{\bf j}\right] n |\tilde{f}_{{\bf j},n}|^2.
\end{eqnarray} 
Here, $\mu_{\bf j}=\mu-\epsilon_{\bf j}$ is the effective chemical potential at the site ${\bf j}$ and 
$\tilde{\Phi}_{\bf j}=\sum_n \sqrt{n}\tilde{f}_{{\bf j},n-1}^{\ast}\tilde{f}_{{\bf j},n}$ is the superfluid order parameter for the steady state.
We will explain how to calculate the excitation spectra from Eq.~(\ref{eq_df/dt}) in the next subsection.

The Gutzwiller approximation has been often applied to the Bose-Hubbard model in order to analyze various phenomena and properties of Bose gases in optical lattices. This approximation is more accurate in higher dimensions because of its mean-field nature. Indeed, thorough comparisons with experiments have shown that it can quantitatively describe several interesting properties in 3D, such as the excitation spectra~\cite{PRL106_205303} and the critical momenta~\cite{PRL99_150604}. The qualitative validity of this approximation is thought to hold also in 2D, given that it correctly captures the order of the SF-to-MI quantum phase transition. In contrast, it is widely known that the Gutzwiller approximation miserably fails in 1D except for very weak interactions $U\ll 1$. Nevertheless, in the present paper we analyze 1D systems using the Gutzwiller approximation because the simplicity of 1D systems is very useful to illustrate basic properties of the superfluid transport that holds in the level of the approximation. 

\subsection{Excitation  spectrum}
Previous works have described prescriptions to calculate the excitation spectra of the homogeneous Bose-Hubbard model within the Gutzwiller approximation~\cite{EPL72_162, PRA84_033602}.
We analyze the excitation spectra not only to investigate the stability of Bose gases in homogeneous systems, but also to obtain collective-mode frequencies in trapped systems.
While Kovrizhin {\it et al.}~have used an extended version of the prescriptions to calculate the excitations in trapped systems~\cite{arXiv0707_2937v1}, they have not explicitly explained their formulations.
In this subsection, we present a detailed explanation of the extended version that allows us to deal with spatially inhomogeneous systems. 
 
We consider a small fluctuation $\delta f_{{\bf j},n}$ around a steady solution $\tilde{f}_{{\bf j},n}$,
\begin{eqnarray}
f_{{\bf j},n}(t) = \left[ \tilde{f}_{{\bf j},n}+\delta f_{{\bf j},n}(t) \right] e^{-i\tilde{\omega}_{\bf j} t},
\label{eq_f+df}
\end{eqnarray} 
We substitute Eq.~(\ref{eq_f+df}) into Eq.~(\ref{eq_df/dt}) and linearize the equation of motion with respect to the fluctuation $\delta f_{{\bf j},n}$. 
The Bogoliubov transformation on the fluctuation,
\begin{eqnarray}
\delta f_{{\bf j}, n}=u_{{\bf j}, n} e^{-i\omega t}+v_{{\bf j},n}^{\ast} e^{i\omega^{\ast} t}
\label{ex_bogo},
\end{eqnarray} 
leads the following linear equations,
\begin{eqnarray}
\omega u_{{\bf j}, n}&=&-\sum_m \sum_{\alpha=1}^d \left[
\left\{ \sqrt{nm} \tilde{f}_{{\bf j},n-1} \tilde{f}_{{\bf j} - {\bf e}_{\alpha}, m-1}^{\ast}
+\sqrt{(n+1)(m+1)}\tilde{f}_{{\bf j}, n+1} \tilde{f}_{{\bf j}-{\bf e}_{\alpha}, m+1}^{\ast}
\right\}
u_{{\bf j} - {\bf e}_{\alpha}, m}
\right. 
\nonumber\\
&& + \left\{
\sqrt{nm}\tilde{f}_{{\bf j}, n-1}\tilde{f}_{{\bf j}+{\bf e}_{\alpha}, m-1}^{\ast}
+\sqrt{(n+1)(m+1)}\tilde{f}_{{\bf j}, n+1}\tilde{f}_{{\bf j}+{\bf e}_{\alpha}, m+1}^{\ast}
\right\}
u_{{\bf j}+{\bf e}_{\alpha}, m}
\nonumber\\
&& + \left\{
\sqrt{n(m+1)}\tilde{f}_{{\bf j}, n-1}\tilde{f}_{{\bf j}-{\bf e}_{\alpha}, m+1}
+\sqrt{(n+1)m}\tilde{f}_{{\bf j}, n+1}\tilde{f}_{{\bf j}-{\bf e}_{\alpha}, m-1}
\right\}
v_{{\bf j} - {\bf e}_{\alpha},m}
\nonumber\\
&& \left. 
+\left\{
\sqrt{n(m+1)}\tilde{f}_{{\bf j}, n-1}\tilde{f}_{{\bf j}+{\bf e}_{\alpha}, m+1}
+\sqrt{(n+1)m}\tilde{f}_{{\bf j}, n+1}\tilde{f}_{{\bf j}+{\bf e}_{\alpha},m-1}
\right\} v_{{\bf j}+{\bf e}_{\alpha}, m}
\right]
\nonumber \\
&& - \sum_{\alpha=1}^d 
\left[\sqrt{n}(\tilde{\Phi}_{{\bf j}-{\bf e}_{\alpha}}+\tilde{\Phi}_{{\bf j}+{\bf e}_{\alpha}})
u_{{\bf j}, n-1} 
+\sqrt{n+1}(\tilde{\Phi}_{{\bf j}-{\bf e}_{\alpha}}^{\ast}+\tilde{\Phi}_{{\bf j}+{\bf e}_{\alpha}}^{\ast})
u_{{\bf j}, n+1}
\right]
\nonumber\\
&& +\left[
\frac{U}{2}n(n-1)-n\mu_{\bf j}-\tilde{\omega}_{\bf j} 
\right]
u_{{\bf j}, n},
\label{eq_ex_u}
\end{eqnarray}
\begin{eqnarray}
-\omega v_{{\bf j}, n}&=&-\sum_m \sum_{\alpha=1}^d \left[
\left\{ \sqrt{nm} \tilde{f}_{{\bf j},n-1}^{\ast} \tilde{f}_{{\bf j} - {\bf e}_{\alpha}, m-1}
+\sqrt{(n+1)(m+1)}\tilde{f}_{{\bf j}, n+1}^{\ast} \tilde{f}_{{\bf j}-{\bf e}_{\alpha}, m+1}
\right\}
v_{{\bf j} - {\bf e}_{\alpha}, m}
\right. 
\nonumber\\
&& + \left\{
\sqrt{nm}\tilde{f}_{{\bf j}, n-1}^{\ast} \tilde{f}_{{\bf j}+{\bf e}_{\alpha}, m-1}
+\sqrt{(n+1)(m+1)}\tilde{f}_{{\bf j}, n+1}^{\ast} \tilde{f}_{{\bf j}+{\bf e}_{\alpha}, m+1}
\right\}
v_{{\bf j}+{\bf e}_{\alpha}, m}
\nonumber\\
&& + \left\{
\sqrt{n(m+1)}\tilde{f}_{{\bf j}, n-1}^{\ast} \tilde{f}_{{\bf j}-{\bf e}_{\alpha}, m+1}^{\ast}
+\sqrt{(n+1)m}\tilde{f}_{{\bf j}, n+1}^{\ast} \tilde{f}_{{\bf j}-{\bf e}_{\alpha}, m-1}^{\ast}
\right\}
u_{{\bf j} - {\bf e}_{\alpha},m}
\nonumber\\
&& \left. 
+\left\{
\sqrt{n(m+1)}\tilde{f}_{{\bf j}, n-1}^{\ast} \tilde{f}_{{\bf j}+{\bf e}_{\alpha}, m+1}^{\ast}
+\sqrt{(n+1)m}\tilde{f}_{{\bf j}, n+1}^{\ast} \tilde{f}_{{\bf j}+{\bf e}_{\alpha},m-1}^{\ast}
\right\} u_{{\bf j}+{\bf e}_{\alpha}, m}
\right]
\nonumber \\
&& - \sum_{\alpha=1}^d 
\left[\sqrt{n}(\tilde{\Phi}_{{\bf j}-{\bf e}_{\alpha}}^{\ast}+\tilde{\Phi}_{{\bf j}+{\bf e}_{\alpha}}^{\ast})
v_{{\bf j}, n-1} 
+\sqrt{n+1}(\tilde{\Phi}_{{\bf j}-{\bf e}_{\alpha}}+\tilde{\Phi}_{{\bf j}+{\bf e}_{\alpha}})v_{{\bf j}, n+1}
\right]
\nonumber\\
&& +\left[
\frac{U}{2}n(n-1)-n\mu_{\bf j}-\tilde{\omega}_{\bf j} 
\right]
v_{{\bf j}, n}.
\label{eq_ex_v}
\end{eqnarray}
Solving these equations, one obtains the frequencies $\omega$ and the wave functions 
$(u_{{\bf j}, n}, v_{{\bf j}, n})$ of excitations, which allow us to discriminate stability of a steady state~\cite{NJP5_104}.
When all the excitations satisfy the condition ${\mathcal N} \omega \geq 0$, the steady state is stable, where ${\mathcal N}=\sum_{\bf j}\sum_n(|u_{{\bf j}, n}|^2 - |v_{{\bf j}, n}|^2)$ is the normalization constant. 
If there exist excitations satisfying ${\mathcal N}\omega < 0$, the state is energetically unstable. This instability is called the Landau instability.
The emergence of excitations with complex frequencies, i.e. $|{\rm Im}[\omega]|>0$, signals the DI which means exponential growth of the fluctuation in time.
In a system of atomic gases at sufficiently low temperatures, while the LI cannot destabilize the system because of the lack of energy-dissipation processes, the DI can drastically break down the system~\cite{PRA72_013603, JPB39_S101, PRA74_053611}.
Hence, it is important to calculate the critical momentum for the DI, even though it is always larger than the Landau critical momentum.

\subsection{Local momentum}
We characterize the transport in trapped systems by the local momentum rather than the center-of-mass (COM) momentum, and compare it with the critical momenta in homogeneous systems.
The local momentum in the direction $\alpha$ is expressed as
\begin{eqnarray}
p_{\bf j}^{\alpha}= \sin^{-1}\left(\frac{I_{\bf j}^{\alpha}}{2\sqrt{n_{\bf j}^{\rm c}n_{{\bf j} +{\bf e}_\alpha}^{\rm c}}}\right),
\label{eq_pj}
\end{eqnarray}
where 
$n_{\bf j}^{\rm c} = |\Phi_{\bf j}|^2$ is the condensate density at the site ${\bf j}$ and $I_{\bf j}^{\alpha}=(\Phi^{\ast}_{\bf j}\Phi_{{\bf j}+{\bf e}_\alpha} - \Phi_{\bf j}\Phi^{\ast}_{{\bf j} +{\bf e}_\alpha})/i$ is the current which flows from the site ${\bf j}$ to the site ${\bf j}+{\bf e}_\alpha$. 
We will show in Sec.~\ref{sec.4} that significant damping of dipole oscillations occurs when the maximum local momentum in a trapped system exceeds the critical momentum for the DI in a homogeneous system.

\section{Critical momenta in homogeneous optical lattices\label{sec.3}}
In this section, we numerically solve Eqs.~(\ref{eq_ex_u}) and (\ref{eq_ex_v}) to obtain the excitation spectra for the homogeneous Bose-Hubbard systems at $\nu = 1$ in 1D, 2D, and 3D, where $\nu$ is the filling factor. Using the discriminant described in the previous section, we determine the critical momenta for the LI and the DI. Notice that the critical momenta for the DI have been obtained in Refs.~\cite{PRL95_020402, PRA71_063613} by solving the equation of motion (\ref{eq_df/dt}) in real-time in a situation where the on-site interaction $U$ or the flow momentum is slowly increased. A clear advantage of our method over the previous work is that the critical momenta for the LI is also available. We also emphasize that one can calculate the critical momenta for the DI more accurately. In the limit of $U\rightarrow 0$, for instance, while the critical momentum obtained in Refs.~\cite{PRL95_020402, PRA71_063613} is a little larger than the prediction by the GP mean-field theory, i.e. $p=\pi/2$~\cite{PRL89_170402}, our method gives a precise agreement with the GP prediction as we will see below.

To calculate the excitation spectra for moving Bose gases, we introduce the phase twist term in Eq.~(\ref{eq_H}) as
\begin{eqnarray}
\hat{a}_{\bf j}^{\dag}\hat{a}_{{\bf j} + {\bf e}_\alpha} \rightarrow \hat{a}_{\bf j}^{\dag}\hat{a}_{{\bf j} + {\bf e}_\alpha}e^{i{\bf p}\cdot{\bf e}_{\alpha}}
\end{eqnarray}
which represents the situation where the optical lattice is moving at a constant momentum $-{\bf p}$.
Forcing the coefficients $f_{{\bf j},n}$ to be real numbers, we carry out imaginary-time evolution of Eq.~(\ref{eq_df/dt}) to obtain the steady state with zero momentum, which is equivalent to a state with momentum ${\bf p}$ in a static optical lattice. Substituting the coefficients $f_{{\bf j},n}$ of the steady state into Eqs.~(\ref{eq_ex_u}) and (\ref{eq_ex_v}) and solving them, we obtain the excitation spectra of the current-carrying state.

\begin{figure}[htbp]
\begin{center}
\begin{tabular}{cc}
\subfigure[]{\includegraphics[height=4cm]{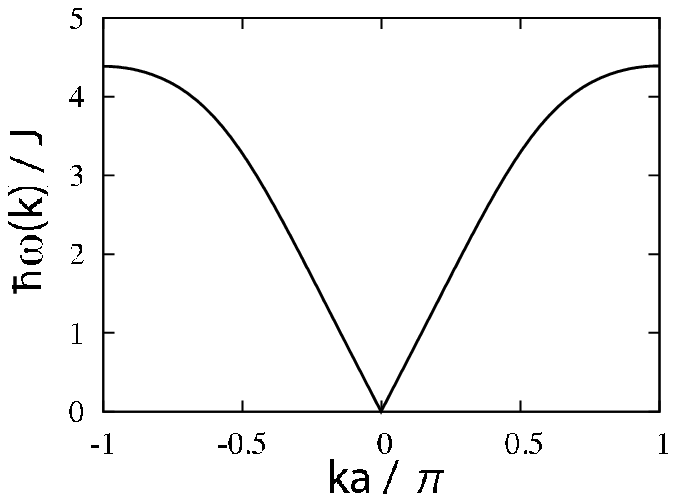}}
\subfigure[]{\includegraphics[height=4cm]{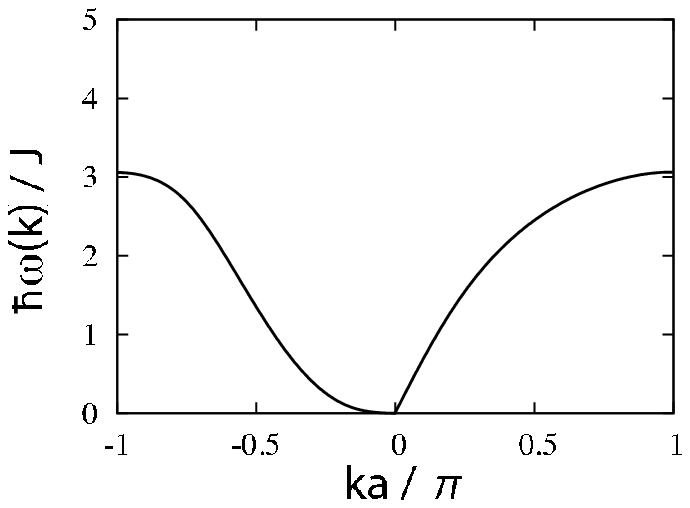}}
\end{tabular}
\begin{tabular}{cc}
\subfigure[]{\includegraphics[height=4cm]{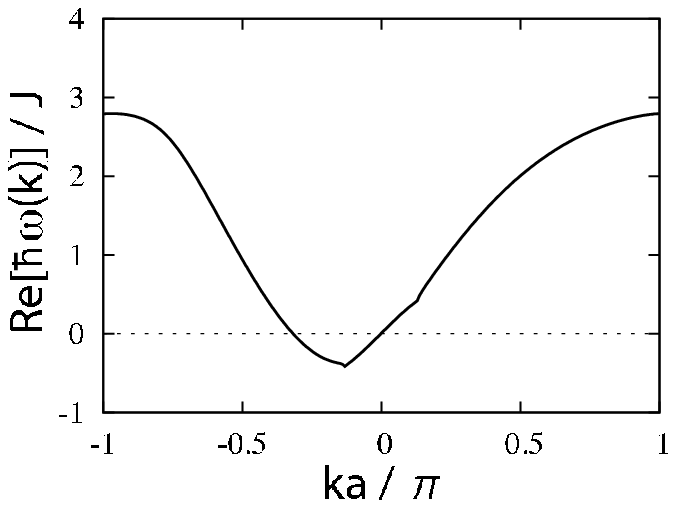}}
\subfigure[]{\includegraphics[height=4cm]{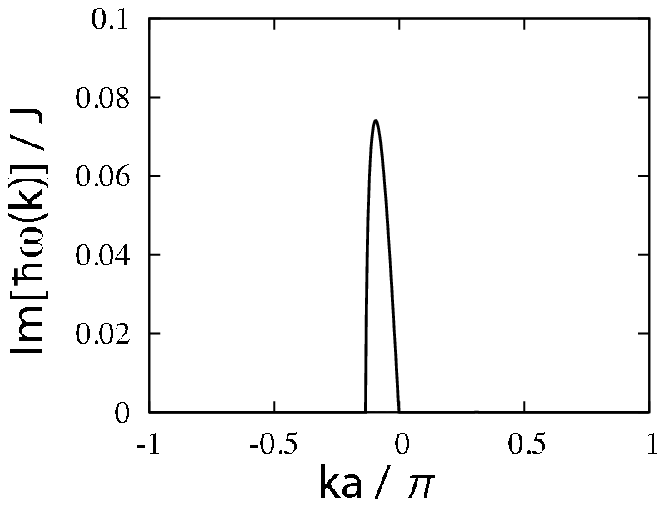}}
\end{tabular}
\caption{\label{fig:homo_ex_u3}
Excitation spectra $\hbar \omega(k)$ in 1D homogeneous optical lattices for $U/J=3$,  $\nu=1$, and different values of $p$.
(a) : $p=0$, (b) : $p=p_{\rm LI} \simeq 0.85$, where the LI sets in, 
(c) and (d) : $p=0.94>p_{\rm DI1}$.}
\end{center}
\end{figure}

In Fig.~{\ref{fig:homo_ex_u3}}, we show the excitation spectra of Bose gases in 1D homogeneous optical lattices for $U=3$ and different values of the flow momentum $p$.  
We plot the first branch of the excitations that corresponds to the well-known Bogoliubov spectrum, because only this branch is relevant to the LI and the DI in most of the parameter regions.
When $p=0$, this branch has a phonon dispersion at $|k| \ll 1$ (Fig.~{\ref{fig:homo_ex_u3}}(a)). When $p$ increases, the slope of the phonon dispersion for $k<0$ decreases, and it reaches zero at $p=p_{\rm LI}\simeq 0.85$ as seen in Fig.~{\ref{fig:homo_ex_u3}}(b). This signals the onset of the LI. 
With $p$ increases further and exceeds a certain threshold, the excitations at $|k|\ll 1$ have a finite imaginary part, which signals the DI. We call this threshold $p_{\rm DI1}$.

\begin{figure}[htbp]
\begin{center}
\begin{tabular}{ccc}
\subfigure[]{\includegraphics[height=4cm]{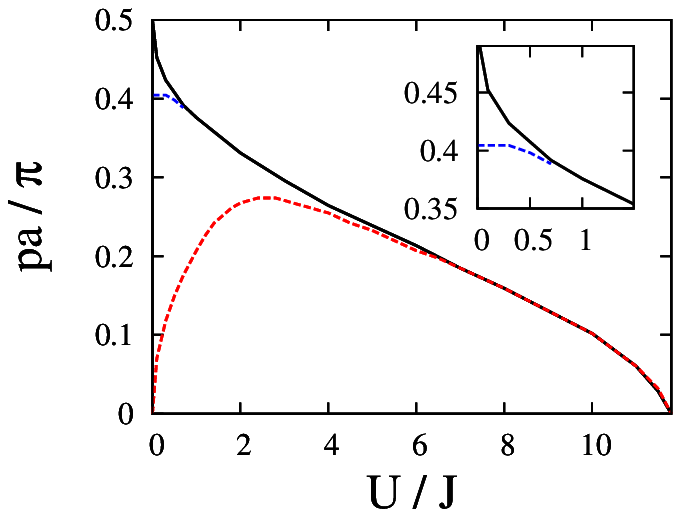}}
\subfigure[]{\includegraphics[height=4cm]{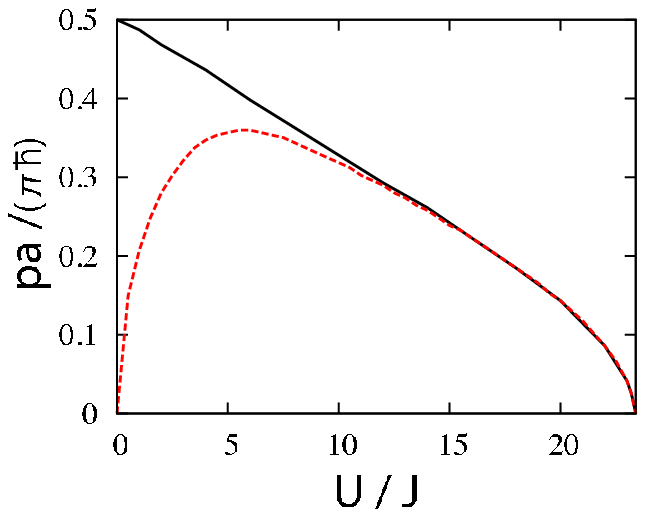}}
\subfigure[]{\includegraphics[height=4cm]{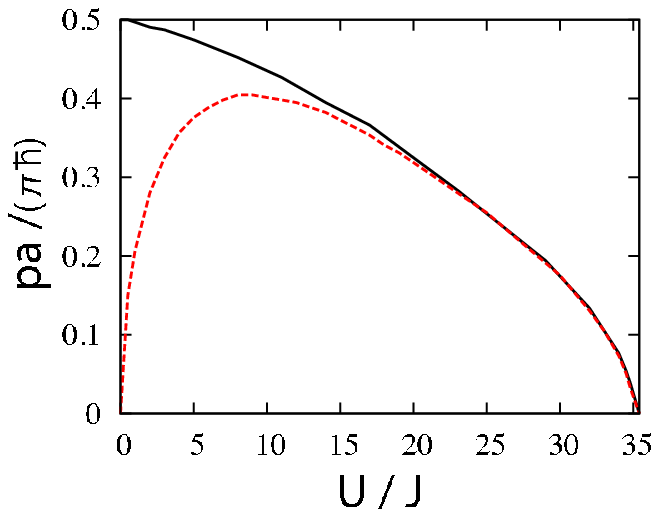}}
\end{tabular}
\caption{\label{fig:homo_critical} (Color online)
Critical momenta versus $U/J$ in homogeneous optical lattices for $\nu=1$.
(a): 1D, (b): 2D, and (c) 3D. 
The dashed red, solid black, and dotted blue lines represent the critical momenta for the LI ($p_{\rm LI}$), the DI caused by excitations with long wavelengths ($p_{\rm DI1}$), and the DI caused by excitations with short wavelengths ($p_{\rm DI2}$).
}
\end{center}
\end{figure}

In Fig.~{\ref{fig:homo_critical}}, we show the critical momenta for the LI (dashed red lines) and the DI caused by excitations with $|k|\ll 1$ (solid black lines) in 1D, 2D, and 3D. 
In any dimensions, $p_{\rm LI}\rightarrow 0$ and $p_{\rm DI1}\rightarrow \pi/2$ when $U\rightarrow 0$.
When $U$ increases, $p_{\rm DI1}$ monotonically decreases and $p_{\rm LI}$ approaches $p_{\rm DI1}$. In the strongly interacting region ($U\gtrsim U_{\rm c}/2$), $p_{\rm LI}$ takes almost the same value as $p_{\rm DI1}$, and both momenta reaches zero at $U=U_{\rm c}$, where $U_{\rm c}$ denotes the SF-MI transition point.
The behaviors of $p_{\rm DI1}$ are consistent with the previous work of Refs.~\cite{PRL95_020402, PRA71_063613}. 

\begin{figure}[htbp]
\begin{center}
\begin{tabular}{ccc}
\subfigure[]{\includegraphics[height=3cm]{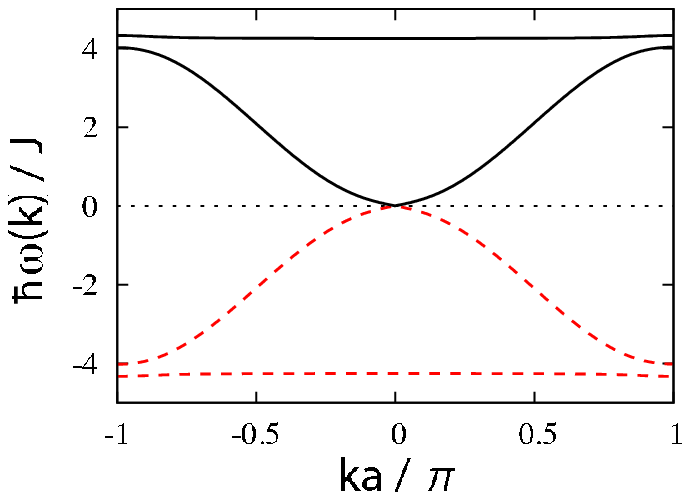}}
\subfigure[]{\includegraphics[height=3cm]{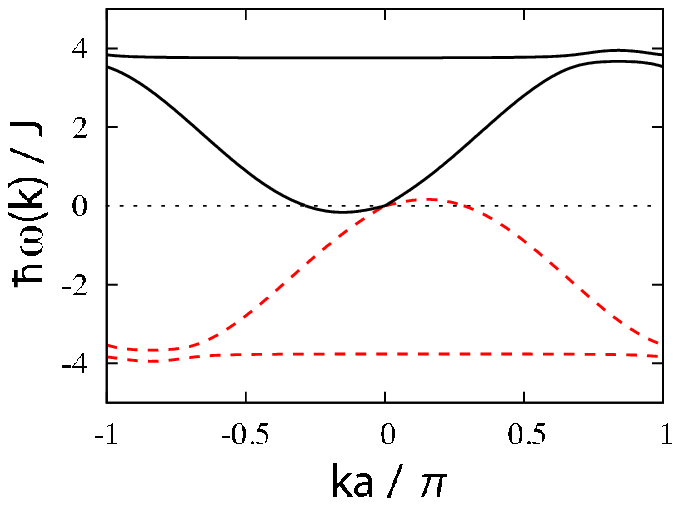}}
\subfigure[]{\includegraphics[height=3cm]{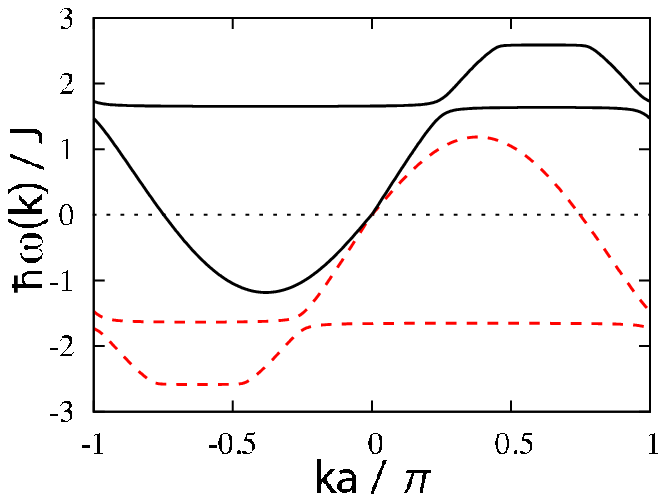}}
\end{tabular}
\begin{tabular}{cc}
\subfigure[]{\includegraphics[height=3cm]{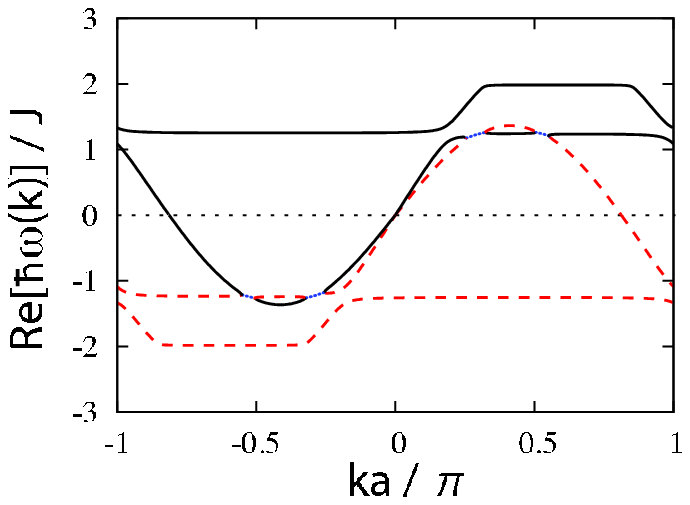}}
\subfigure[]{\includegraphics[height=3cm]{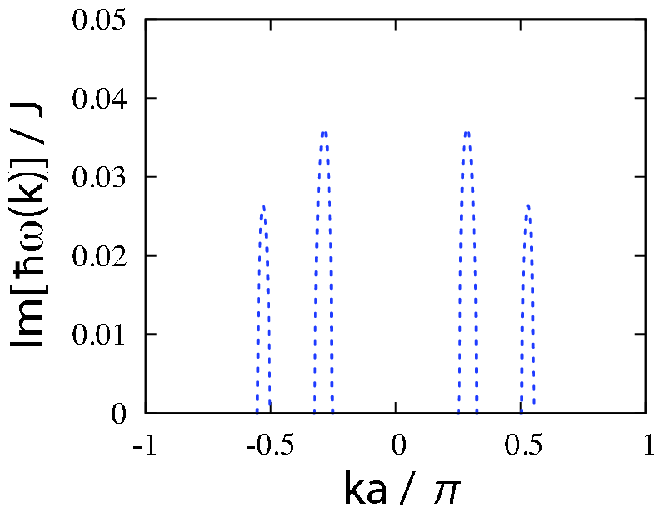}}
\end{tabular}
\caption{\label{fig:homo_ex_u0,1} 
Excitation spectra $\hbar \omega(k)$ in 1D homogeneous optical lattices 
for the weak interaction $U/J=0.1$, $\nu = 1$, and different values of $p$. (a): $p=0$, (b): $p=0.5$, 
where the avoided crossing of the Bogoliubov and amplitude modes starts to occur, (c): $p=1.2$, 
(d) and (e): $p=1.3$, where excitations with short wavelengths causes the DI.
The solid black, red dashed, and blue dotted lines represent the normal modes, the anti-modes, and the modes whose imaginary part is finite.
}
\end{center}
\end{figure}

When the interaction is sufficiently weak in 1D, a DI caused by excitations with short wavelengths precedes the DI caused by excitations with $|k|\ll 1$. The critical momentum for this DI $p_{\rm DI2}$ is plotted by the dotted blue line in Fig.~\ref{fig:homo_critical}(a).
To understand this instability, we show in Fig.~\ref{fig:homo_ex_u0,1} the excitation spectra including the Bogoliubov mode, the amplitude mode, and the anti-modes of the two modes~\cite{footnote1} for $U=0.1$ and different values of $p$.  
When $p$ increases, the amplitude mode declines and exhibits an avoided crossing with the Bogoliubov mode as seen in Figs.~\ref{fig:homo_ex_u0,1}(b) and (c).
With increasing $p$ further, the amplitude (Bogoliubov) mode is coupled with the anti-Bogoliubov (anti-amplitude) mode and these coupled modes cause the DI~\cite{NJP5_104}.  
We suggest that this DI leads to the formation of density waves with ordering vectors $k \simeq \pm 0.3\pi$ and $\pm 0.5 \pi$. While this type of DI has been previously found in the presence of off-site interactions~\cite{PRA82_043625, PRL100_255301, PRA80_043612, PRA80_063627}, this is the first example of such a DI in the standard Bose-Hubbard model only with the on-site interaction.

\section{Dipole oscillations in the standard Bose-Hubbard system\label{sec.4}}
In this section, we consider systems of Bose gases in 1D and 2D optical lattices combined with a parabolic trap. 
We fix the total number of particles to be $N=45$ in 1D and $N=3000$ in 2D such that the Mott plateau at unit filling forms in the central region of the trap when $U$ exceeds $U_{\rm c}$.  
We first calculate the ground state via imaginary-time evolution of Eq.~(\ref{eq_df/dt}).
We next calculate real-time dynamics subjected to a sudden displacement of the trap center that induces a dipole oscillation.
Notice that while similar dynamics have been studied in Refs.~\cite{arXiv0707_2937v1, PRA76_051603} using the same Gutzwiller approximation, the previous studies did not address the relation between the dipole oscillations and the critical momenta, which is our main interest here.

\subsection{Stability and damping of dipole oscillations}
\begin{figure}[htbp]
\begin{center}
\begin{tabular}{ccc}
\subfigure[]{\includegraphics[height=4cm]{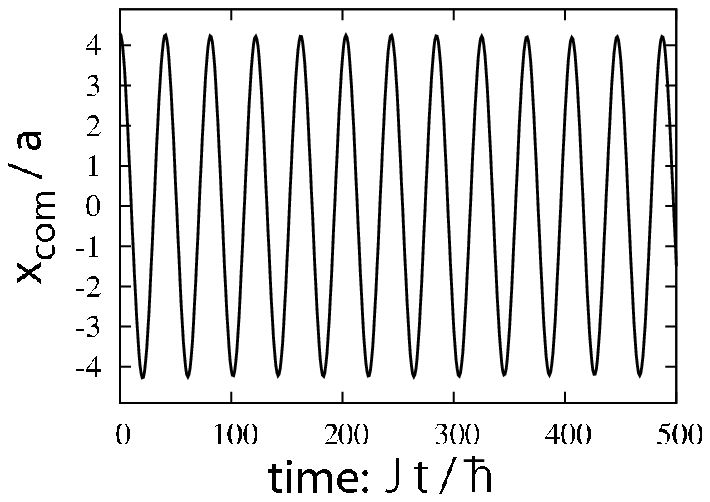}}
\subfigure[]{\includegraphics[height=4cm]{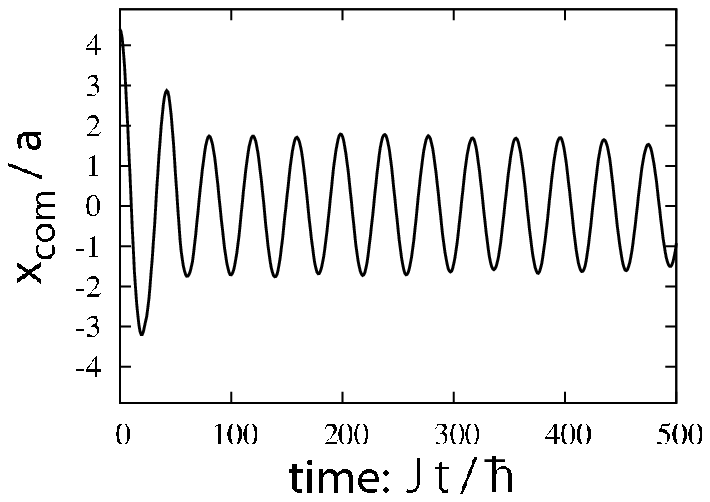}}
\end{tabular}
\caption{\label{fig:trap_com_u6} 
Time evolution of the COM $x_{\rm com}(t)$ in 1D optical lattices combined with the paraboric trap for $U/J=6$. 
(a): stable dipole oscillation at $D=4.3$, 
(b): damped dipole oscillation at $D=4.4$.}
\end{center}
\end{figure}
To illustrate basic properties of the dynamics of dipole oscillations, we start our analyses with the 1D case. 
In Fig.~\ref{fig:trap_com_u6}, we show typical examples of the COM motion for stable and damped dipole oscillations, where the COM is given by $x_{\rm com}=\sum_{j} j n_{j} / \sum_j n_j$.
The amplitude of the dipole oscillation hardly changes as long as the displacement $D$ is smaller than a certain critical value $D_{\rm c}$ as seen in Fig.~\ref{fig:trap_com_u6}(a).
This dissipationless motion is a clear characteristic of superflow.
In contrast, the oscillation is remarkably damped when $D_{\rm c}$ is exceeded (see Fig.~\ref{fig:trap_com_u6}(b)). 
It is worth noting that the dynamics for these two cases can be clearly distinguished because the transition to the damped oscillation occurs very sharply. 
Thanks to this property, one can determine the critical displacement accurately.

\subsection{Local momentum}
\begin{figure}[htbp]
\begin{center}
\begin{tabular}{ccc}
\subfigure[]{\includegraphics[height=4cm]{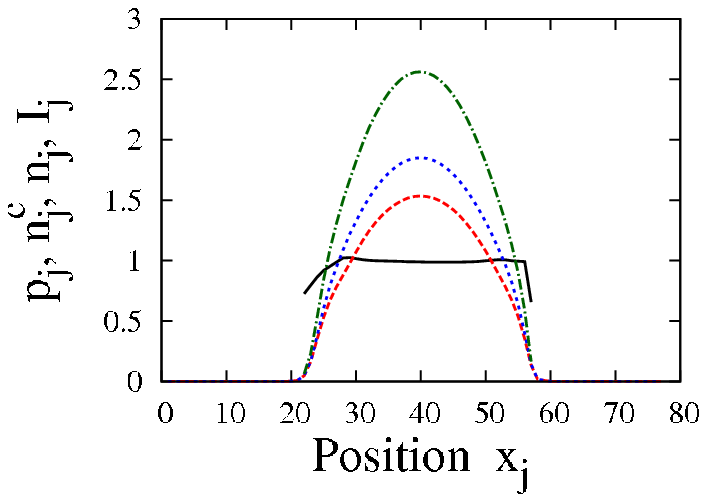}}
\subfigure[]{\includegraphics[height=4cm]{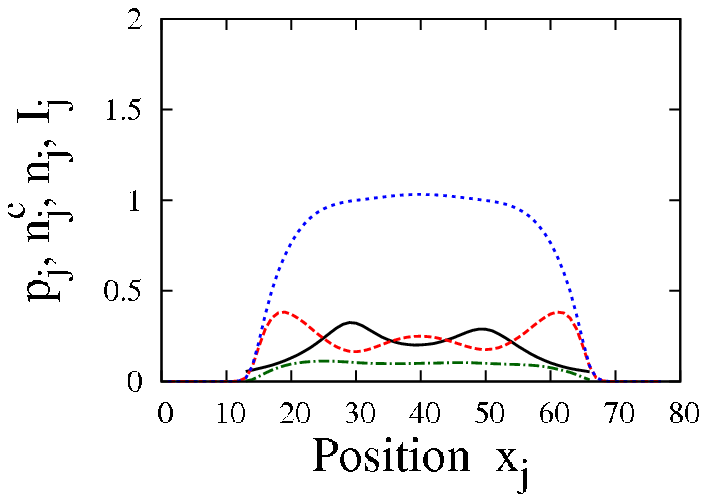}}
\end{tabular}
\caption{\label{fig:trap_local} (Color online) 
Local momentum $p_j$ (solid black line), 
local condensate density $n^{\rm c}_{j}=|\Phi_j|^2$ (dashed red line), local density $n_j$  (dotted black line), local current $I_j$ (dot-dashed green line) in 1D at $D=D_{\rm c}$ and $t=t_{\rm max}$, where $U/J=2$ (a) and $U/J=10$ (b).}
\end{center}
\end{figure} 
For comparison with the critical momenta in the homogeneous system, we calculate two types of momentum in the trapped system, namely the COM momentum, $p_{\rm com}=\sin^{-1} (v_{\rm com}/2)$, and the local momentum defined in Eq.~(\ref{eq_pj}), where $v_{\rm com}$ is the COM velocity.
In Fig.~\ref{fig:trap_local}, we show the snap shots of the local momentum $p_j$, the condensate density $n_{j}^{\rm c}\equiv |\Phi_j|^2$, and the density $n_j$ at the critical displacement $D=D_{\rm c}$ for a weak interaction $U=2$ (a) and a strong interaction $U=10$ (b).
The time is set such that the momentum of COM takes the maximum value during the time evolution.
We define this time as $t_{\rm max}$.
For the weak interaction, the local momentum is almost constant. 
On the other hand, for the strong interaction in Fig.~\ref{fig:trap_local}(b), the local momentum is significantly dependent on the position so that it peaks at the unit-filling points.
This happens because the condensate density in the regions close to unit filling is strongly suppressed as a precursor of the formation of the Mott plateaus. 
Due to the strong spatial dependence of the local momentum, the COM momentum is noticeably smaller than the maximum local momentum, especially near the Mott transition.

\subsection{Critical momentum}
\begin{figure}[htbp]
\begin{center}
\begin{tabular}{cc}
\subfigure[]{\includegraphics[height=4cm]{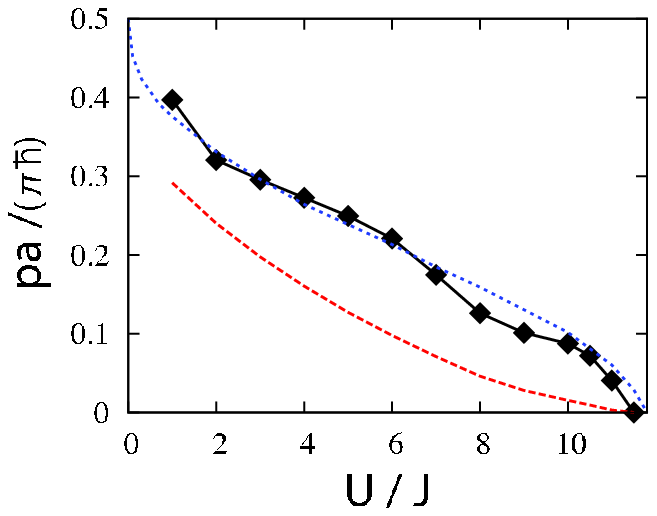}}
\subfigure[]{\includegraphics[height=4cm]{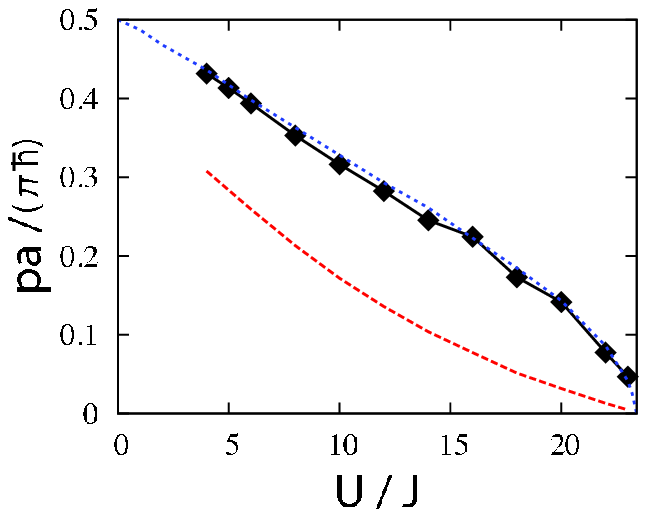}}
\end{tabular}
\caption{\label{fig:trap_critical} (Color online)
Maximum local momentum $p_{\rm max}$ (solid black lines with symbols) and the COM momentum $p_{\rm com}$ (dashed red lines) at $D=D_c$ and $t=t_{\rm max}$ as functions of $U/J$. 
(a) : 1D and (b) : 2D. 
The dotted blue lines represent $p_{\rm DI1}$ in the homogeneous lattice systems, which are plotted in Figs.~\ref{fig:homo_critical}(a) and (b) with the solid black lines.}
\end{center}
\end{figure}

In Fig.~\ref{fig:trap_critical}, we show the maximum local momentum $p_{\rm max}$ and the COM momentum $p_{\rm com}$ at $D=D_{\rm c}$ and $t=t_{\rm max}$ as functions of $U$.
For comparison, we also plot the critical momenta $p_{\rm DI1}$ obtained in the homogeneous systems at unit filling.
We find that $p_{\rm max}(t=t_{\rm max})$ in the trapped system quantitatively agrees with $p_{\rm DI1}$. 
In contrast, $p_{\rm com}(t=t_{\rm max})$ drastically deviates from $p_{\rm DI1}$. 
These results lead to the important conclusion; for detecting the critical momenta for the homogeneous lattice systems from the dipole oscillations, one needs to measure the local momentum rather than the COM momentum that can be measured relatively easily from the time-of-flight images~\cite{PRL94_120403, nature2008, PRL86_4447}. 
While the local momentum has never been experimentally measured so far, it should be available in future experiments because recent experiments have developed techniques to address local observables at the scale of a single lattice site~\cite{science2012, science2011}.

\subsection{Mode coupling}
\begin{figure}[h!]
\includegraphics[height=8cm]{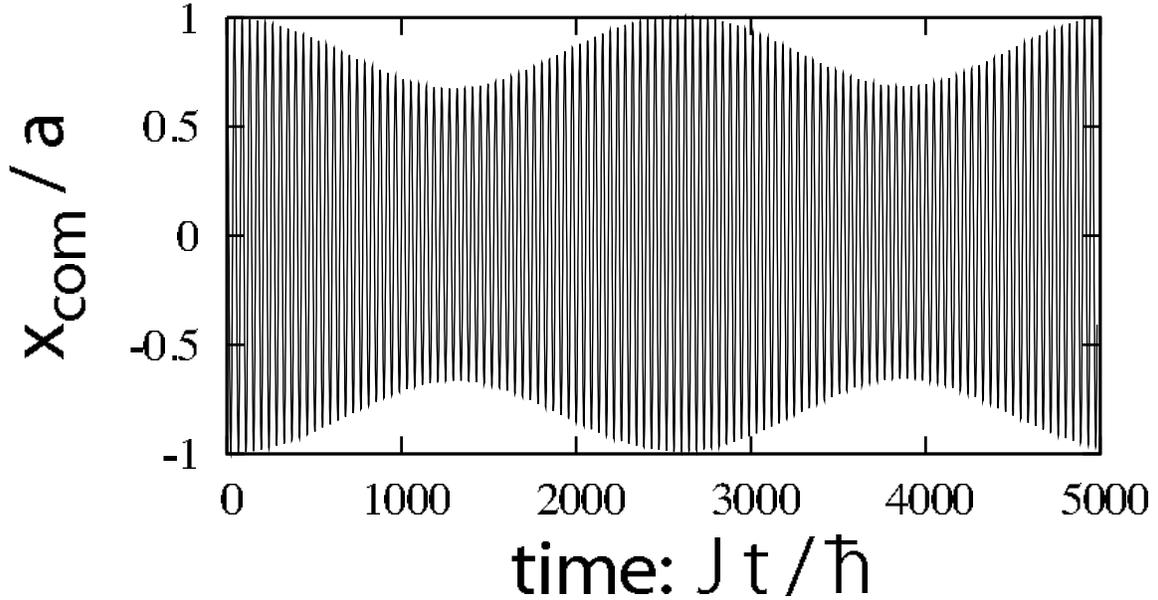}
\caption{\label{fig:mode_com_u7,78} 
The motion of the COM $x_{\rm com}(t)$ with damping and revival in 1D optical lattices combined with a parabolic trap for $U/J=7.78$.}
\end{figure}
\begin{figure}[htbp]
\begin{center}
\begin{tabular}{ccc}
\subfigure[]{\includegraphics[height=4cm]{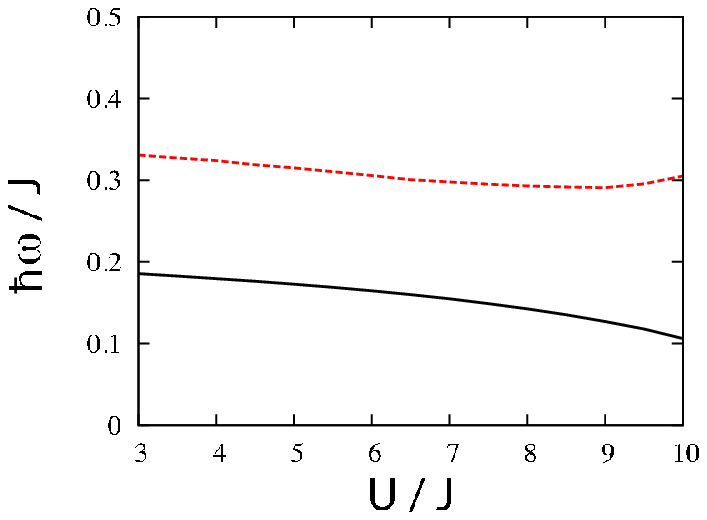}}
\subfigure[]{\includegraphics[height=4cm]{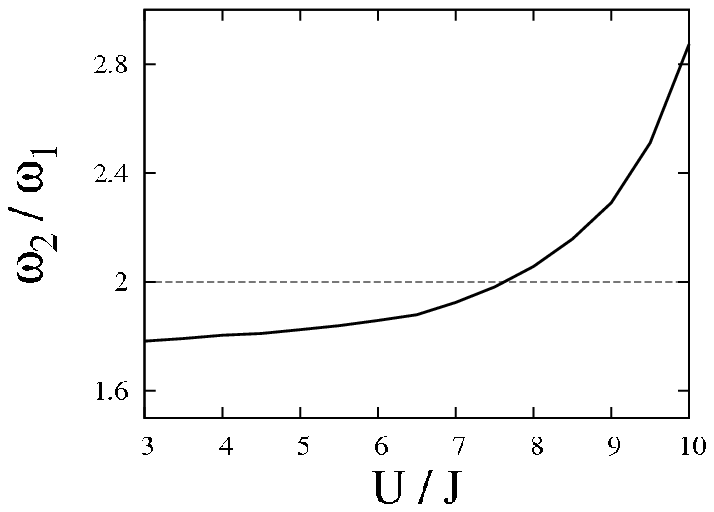}}
\end{tabular}
\caption{\label{fig:mode_bre_dip_ratio}(Color online) 
(a) Frequencies of the first mode $\omega_1$ (solid black line) and the second mode $\omega_2$ (dashed red line) in 1D lattices combined with a parabolic trap. 
(b) Ratio $\omega_2/\omega_1$.}
\end{center}
\end{figure}
We now focus on long-time dynamics of the dipole oscillation.
In this case, we find that the dipole oscillation with the damping and the revival occurs below the critical momentum in a particular range at the interaction $U=7.5 - 7.8$. 
Similar behaviors have been found in Refs.~\cite{arXiv0707_2937v1, PRA76_051603}.  
In Fig.~\ref{fig:mode_com_u7,78}, we show this unusual oscillation for $U =7.78$.
For understanding this phenomenon, we calculate the excitation spectrum by diagonalizing Eqs.~(\ref{eq_ex_u}) and (\ref{eq_ex_v}). 
In Fig.~\ref{fig:mode_bre_dip_ratio}(a), we plot the frequencies of the first and second modes, which correspond to the dipole mode and the breathing mode, respectively. 
We also show the ratio of  the breathing-mode frequency $\omega_2$ to the dipole-mode frequency $\omega_1$ in Fig.~\ref{fig:mode_bre_dip_ratio}(b).
We find that the ratio $\omega_2 / \omega_1$ is nearly equal to two in the regime where the damping and the revival occur. 
Therefore, this phenomenon can be interpreted as indicating that dipole and breathing modes are coupled by the nonlinear effect in Eq.~(\ref{eq_df/dt}). 
In this regime, although the amplitude of the dipole oscillation decreases due to this mode coupling even below the critical momentum, we can determine the critical momentum because the damping caused by mode coupling is much smaller than that caused by the DI.

\section{Dipolar hardcore bosons\label{sec.5}}
In this section, we consider dipolar bosons confined in 2D optical lattices combined with a parabolic trapping potential. 
For simplicity, we consider the hardcore limit ($U\rightarrow \infty$), in which the local Hilbert space is spanned only with $|0 \rangle$ and $|1 \rangle$, and assume that the dipoles are polarized to the direction perpendicular to the lattice plane.
This system is well described by the following Bose-Hubbard model,
\begin{eqnarray}
\hat{H}=-J\sum_{\bf j}\sum_{\alpha=1}^d (\hat{a}^{\dag}_{\bf j} \hat{a}_{{\bf j} + {\bf e}_{\alpha}}+h.c.)+\frac{V}{2}\sum_{{\bf j}\neq {\bf l}} \frac{\hat{n}_{\bf j}\hat{n}_{\bf l}}{r^3_{{\bf j}, {\bf l}}}+\sum_{\bf j} (\epsilon_{\bf j} - \mu) \hat{n}_{\bf j},\label{eq_H_dip}
\end{eqnarray}
where $V$ is the strength of dipole-dipole interaction, $r_{{\bf j}, {\bf l}}=a|{\bf j} - {\bf l}|$ is the distance between the sites ${\bf j}$ and ${\bf l}$. 
Since the dipolar interaction potential decays as $\sim r_{{\bf j}, {\bf l}}^{-3}$, we only include the dipolar interactions in the range  $r_{{\bf j}, {\bf l}}\leq 7a$. Introducing this cutoff merely changes some quantitative natures of the system very slightly, but does not affect any qualitative natures, such as the quantum phase diagram and the critical momentum, when $V \lesssim 10$. 
We analyze this system by using the Gutzwiller approximation as in the previous section.
We again adopt the units of $\hbar = J = a =1$.

\subsection{Ground state}
\begin{figure}[h!]
\includegraphics[height=4cm]{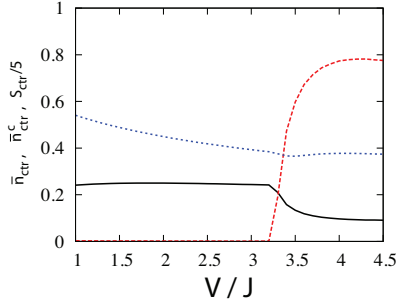}
\caption{\label{fig:dip_gs_1} (Color online) 
The average density $\bar{n}_{\rm ctr}$ (dotted blue line), The average condensate density $\bar{n}^{\rm c}_{\rm ctr}$ (solid black line), and the static structure factor $S_{\rm ctr}(\pi, \pi)$ devided by 5 (dashed red line) around the trap center in the ground state of Eq.~(\ref{eq_H_dip}) are plotted as functions of $V/J$, where $\Omega = 0.01 J$ and $N=900$. 
}
\end{figure}
Let us calculate the ground state of Eq.~(\ref{eq_H_dip}).
We fix the curvature of the parabolic potential and the total number of particles to be $\Omega=0.01$ and $N=900$. 
In this case, the ground state can be the following two types of state as long as $V<4.5$.
For small $V$, the dipolar gas in the ground state consists only of a SF region.
When $V$ is sufficiently large, the ground state has a region of the SS with the checkerboard density wave order around the trap center, and this SS region is surrounded by a SF region. 
As a metaphor, one may imagine a sunny-side up whose yolk and white correspond to the SS and SF regions.  
In order to identify the transition between the two states, we show in Fig.~\ref{fig:dip_gs_1} the average condensate density and the static structure factor around the center of the trap, which are defined as $n^{\rm c}_{\rm ctr} =\sum_{{\bf j} \in {\rm ctr}} |\Phi_{\bf j}|^2/25$ 
and $S_{\rm ctr}(\pi,\pi)=\sum_{{\bf j}, {\bf l} \in {\rm ctr}}e^{i(\pi, \pi)\cdot ({\bf j}-{\bf l})}\langle n_jn_l\rangle /25$. 
Here $\sum_{\in {\rm ctr}}$ denotes the summation of the $5\times 5$ sites around the center of the trap. 
For comparison with the results in Ref.~\cite{PRA82_043625}, we also plot the average density $\bar{n}_{\rm ctr}=(n_{\bf 0}+n_{{\bf e}_1})/2$.
By using these quantities, we identify the state with $n^{\rm c}_{\rm ctr}\neq 0$ and $S_{\rm ctr}(\pi, \pi)=0$ as the SF state and the state with $n^{\rm c}_{\rm ctr}\neq 0$ and $S_{\rm ctr}(\pi, \pi)\neq 0$ as the SS state.
We find that the boundary of the SF and the SS is $V=3.26$.

\subsection{Critical momentum}
\begin{figure}[htbp]
\begin{center}
\begin{tabular}{ccc}
\subfigure[]{\includegraphics[height=4cm]{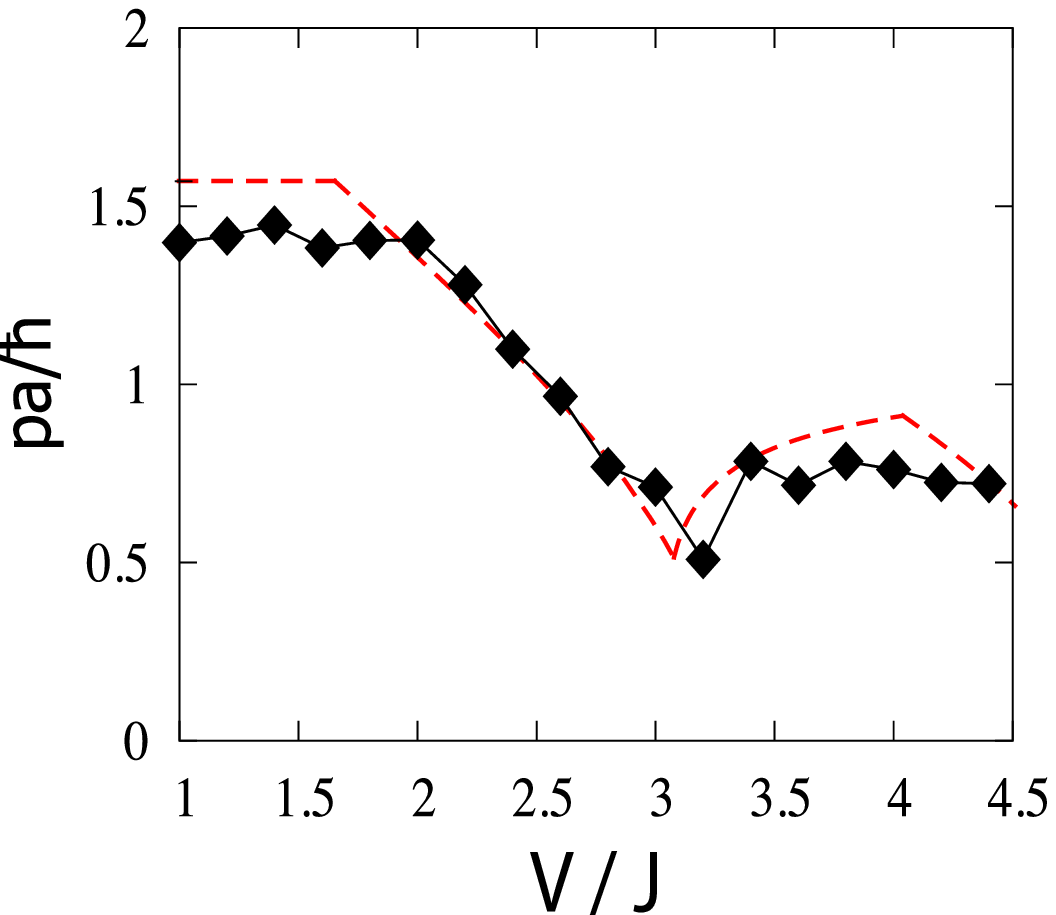}}
\subfigure[]{\includegraphics[height=4cm]{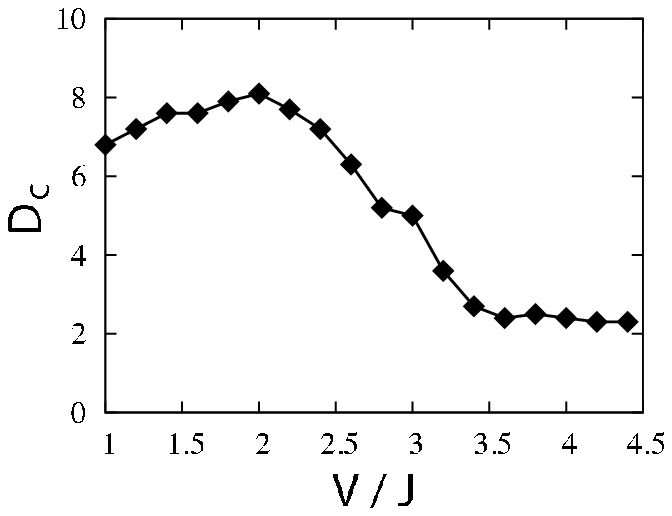}}
\end{tabular}
\caption{\label{fig:g_dip_critical_mc} 
(a) The solid line with symbols represents the maximum local momentum $p_{\rm max}$ at $D=D_{\rm c}$ and $t=t_{\rm max}$ in the system of Eq.~(\ref{eq_H_dip}) as a function of $V/J$.  The dashed red line represents the critical momentum for the DI in the homogeneous case at $\nu=0.4$~\cite{PRA82_043625}.
(b) Critical displacement $D_{\rm c}$ as a function of $V/J$. }
\end{center}
\end{figure}

Having obtained the ground state of Eq.~(\ref{eq_H_dip}), we next calculate the dipole-oscillation dynamics following the way illustrated in the previous section.
We find that the critical displacement $D_{\rm c}$ can be accurately determined also in the case of dipolar hardcore bosons.
By the solid black line with symbols in Fig.~\ref{fig:g_dip_critical_mc}(a), we show the maximum local momentum $p_{\rm max}$ at $D=D_{\rm c}$ and $t=t_{\rm max}$ as a function of $V$.
We clearly see that $p_{\rm max}$ agrees well with the critical momentum for the DI in the homogeneous lattice system at $\nu = 0.4$ obtained in Ref.~\cite{PRA82_043625}. $p_{\rm max}$ takes a distinct minimum at a point close to the transition to the SS state and is significantly smaller  in the SS state than in the SF state. We emphasize that the agreement with the homogeneous case is found in the SS state even though there also exists the SF region with a different critical momentum. This can be attributed to the fact that the critical momentum in the SS phase is smaller than that in the SF phase and that the maximum local momentum is always taken at the center of the SS region where the local condensate density is smallest.

Let us discuss the feasibility of identifying the SS state from the critical momentum.
If one can measure the local momentum, the above-mentioned properties of the critical momentum in the SS state allow for distinguishing the SS state from other possible states, such as the SF, MI, and density-wave insulating states.
However, since it should be easier to measure the critical displacement rather than the critical local momentum, we show the critical displacement as a function of $V$ in Fig.~\ref{fig:g_dip_critical_mc}(b), where the significant reduction of $D_{\rm c}$ in the SS state is clearly seen.
Thus, we suggest that it is possible to identify the SS state by measuring the critical displacement $D_{\rm c}$.

\subsection{Dynamical transition}
\begin{figure}[htbp]
\begin{center}
\begin{tabular}{ccc}
\subfigure[]{\includegraphics[height=6cm]{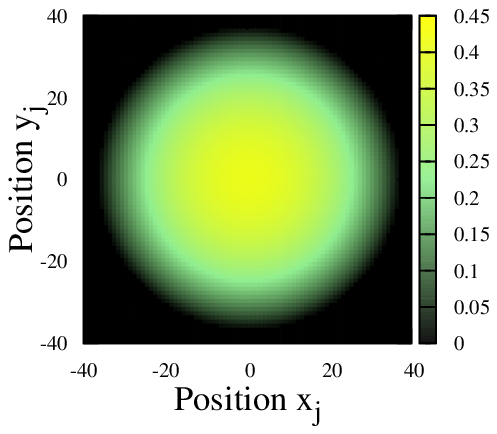}}
\subfigure[]{\includegraphics[height=6cm]{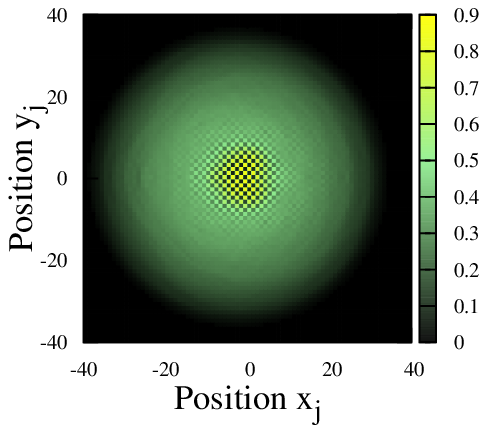}}
\end{tabular}
\caption{\label{fig:f_dip_sf_ss} (Color online) 
Snap shots of the density distribution $n_{\bf j}$ during the dipole-oscillation dynamics of Eq.~(\ref{eq_H_dip}) with $V/J=3.2$, where $Jt/\hbar =0$ (a) and $Jt/\hbar=310$ (b).}
\end{center}
\end{figure}
\begin{figure}[h!]
\includegraphics[height=4cm]{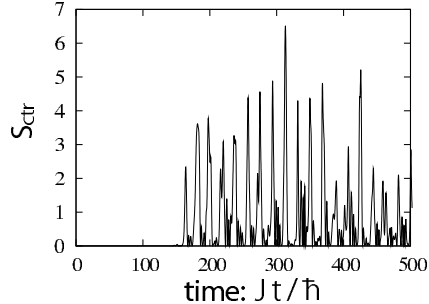}
\caption{\label{fig:g_dip_dy_transition} 
Time evolution of the structure factor $S(\pi, \pi)$ during the dipole oscillation, where $V/J=3.2$ and $D=2$. }
\end{figure}
It has been predicted for the homogeneous case in Ref.~\cite{PRA82_043625} that when the superflow momentum increases in the SF state near the phase boundary to the SS state, the transition to the SS phase can occur due to the DI caused by excitations with ${\bf k}=(\pi,\pi)$.
Since typical experimental setup includes a parabolic trapping potential, 
it is important to confirm whether this dynamical transition induced by superflow occurs or not in trapped systems.
For this purpose, we choose $V=3.2$ such that the ground state is in the SF state close to the transition point.
In Fig.~\ref{fig:f_dip_sf_ss}(a), we show the density profile of the ground state.
Taking this state as the initial state, we compute the dipole-oscillation dynamics of Eq.~(\ref{eq_H_dip}) setting $D=2$.
To quantify the formation of the density wave order associated with the dynamical transition to the SS, we compute the time evolution of the static structure factor $S_{\rm ctr}(\pi, \pi)$, which is shown in Fig.~\ref{fig:g_dip_dy_transition}.
$S_{\rm ctr}(\pi, \pi)$ exhibits repetitive growth and collapse after the time $t \simeq 150$.
In Fig.~\ref{fig:f_dip_sf_ss}(b), we show the density distribution at $t=310$, where the growth of the $S_{\rm ctr}(\pi, \pi)$ is the most prominent within the time scale plotted in Fig.~\ref{fig:g_dip_dy_transition}.
There we clearly see that the checkerboard density wave forms around the trap center.
Thus, the dynamical transition to the SS can occur in the dipole-oscillation dynamics in the trapped system.

\section{Summary\label{sec.6}}
We have studied the superfluid transport of Bose gases in optical lattices by using the Gutzwiller approximation.
In one-dimensional (1D), 2D, and 3D homogeneous systems, we determined the critical momenta for the Landau instability (LI) and the dynamical instability (DI) from the excitation spectra.
Especially, we have found a DI caused by excitations with short wavelengths  in 1D when the on-site interaction is very small. 
In trapped system, we have analyzed the dynamics of dipole oscillations induced by suddenly displacing the trap center.
We have found that the critical momentum defined by the maximum local momentum
in the trap system quantitatively agrees with that in the homogeneous system at unit filling. 
We have also shown that a resonance phenomenon is caused by the coupling between the dipole-oscillation mode and the breathing mode.
Moreover, we have investigated the critical momentum of dipolar hardcore bosons and shown that the critical momentum of the SS state is smaller than that of the SF state as in the case of the homogeneous system.
This result allow us to suggest that the SS state should be identified by measuring the critical displacement.
We finally showed that the dynamical transition from SF state to SS state can be induced by dipole oscillations.

\begin{acknowledgments}
The authors thank A. Polkovnikov, S. Tsuchiya, and D. Yamamoto for useful comments and discussions.
\end{acknowledgments}


\begin{thebibliography}{1}
\bibitem{footnote1}
It can be proven that if Eqs.~(\ref{eq_ex_u}) and (\ref{eq_ex_v}) have a solution with a frequency $\omega$ and ${\mathcal N} > 0$, they also have a solution with a frequency $-\omega$ and ${\mathcal N} < 0$, which we call the anti-mode.
\end{thebibliography}
\end{document}